# Utilization of Virtual Reality Visualizations on Heavy Mobile Crane Planning for Modular Construction

**Navid Kayhani, Hosein Taghaddos, Mojtaba Noghabaee, and Ulrich (Rick) Hermann**

[a] Research Assistant at TECNOSA R&D Center, M.Sc. in Construction Engineering and Management, School of Civil Engineering, College of Engineering, University of Tehran, Iran.
[b] Assistant Professor, School of Civil Engineering, College of Engineering, University of Tehran, Iran.
[c] Ph.D. Student, Dept. of Civil, Construction, and Environmental Engineering, North Carolina State University , Raleigh, NC 27695.
[d] Manager of Construction Engineering, PCL Industrial Management Inc., Edmonton, AB, Canada, T6E 3P4
E-mail: navid_kayhani@ut.ac.ir, htaghaddoss@ut.ac.ir, snoghab@ncsu.edu, rhhermann@pcl.com

**Abstract –**
Many kinds of industrial projects involve the use of prefabricated modules built offsite, and installation on-site using mobile cranes. Due to their costly operation and safety concerns, utilization of such heavy lift mobile cranes requires a precise heavy lift planning. Traditional heavy lift path planning methods on congested industrial job sites are ineffective, time-consuming and non-precise in many cases, whereas computer-based simulation models and visualization can be a substantial improving tool. This paper provides a Virtual Reality (VR) environment in which the user can experience lifting process in an immerse virtual environment. Providing such a VR model not only facilitates planning for critical lifts (e.g. modules, heavy vessels), but also it provides a training environment to enhance safe climate prior to the actual lift. The developed VR model is implemented successfully on an actual construction site of a petrochemical plant on a modular basis in which heavy lift mobile cranes are employed.

## 1 Introduction

Effective management and planning of resources are pivotal tasks in construction projects management. In recent decades, many industrial projects are constructed using industrialization/ prefabricated offsite construction methods, particularly modularization methods. Modules (e.g. pipe racks or vessels) are offsite prefabricated elements that assembled in module yards and shipped to the site for installation [1]. A set of modules constructs a larger structure. During the installation process, heavy mobile cranes are employed as the main lifting machinery. This heavy and large construction equipment is highly expensive and required precise planning and control due to safety concerns and the extensive impact on the project schedule. For example, the rental cost of a heavy lift mobile crane may reach $1,500 per hour [2]. Moreover, not only these resources force highly expenses (e.g. mobile crane rental cost, skilled crew, and mat cost) to a project, but also they are just available for a restricted period of time.

Proper use of these heavy mobile cranes can reduce costs and make the construction process faster. In contrast, inappropriate crane operation can lead to overruns, delays, and safety issues. Heavy lift path planning is one of the most significant elements in heavy mobile crane planning and management. Traditional heavy lift path planning methods on congested industrial job sites are ineffective, time-consuming and non-precise in many cases, whereas computer-based simulation and visualization can be considerable improving tools.

With tremendous development of graphical computer aid modeling and also novel interactive hardware, a definite opportunity is provided in architecture engineering and construction (AEC) industry to facilitate the training process and effective learning in virtual environments (i.e. virtual reality, augmented reality and mixed reality) through that. A lot of research has been done in this area, and scholars tried to examine the effectiveness of the application in the construction industry. Additionally, there are also many untapped opportunities in the construction planning process.

The objective of this research is to provide a Virtual Reality (VR) environment in which one can experience a simulated lifting process in an immerse virtual environment. Providing such a VR model not only facilitates planning for blind or critical lifts (e.g. modules, heavy vessels) but also provides a safe training environment prior to the actual lift. Moreover, using VR technology make it possible to provide useful extra information, and by so doing, operators, planners, and engineers will have the opportunity to obtain a common understanding and would be capable to cooperate to improve the plan.

## 2  Related Works

Throughout the recent decades, computers have played a significant role in construction. Scheduling and many other planning tasks are done with the support of computers. In the recent years, other applications have been also investigated particularly in construction operation visualization. Visualization aims to verify trustworthiness and to provide presentable insights about the modeled system. For instance, four-dimensional (4D) computer-supported design CAD technology presents a visualization of the advancement of construction by linking three-dimensional (3D) computer models of the project and construction schedules [3]. 4D models barely enable users to envision a construction schedule previously developed. However, they are not able to simulate the construction activities [4]. More recently, a faster design process and more optimized end products. The first industrial use of VR was introduced when virtual prototypes started to be used in the design process. VR is a virtual prototyping technology that provides a three-dimensional scene that can be manipulated in real time and used collaboratively to explore and navigate the 3D model. By simulation of the construction process, VR makes users to get a better understanding and perception about the actual environment. As illustrated in Figure 1, there are many potential application areas for VR systems [5]. Various conducted research in the case of VR technology applications has confirmed that VR can bring vital value improvement and cost declines to the construction process [6].

Management of heavy construction equipment is a sophisticated task due to highly dynamic activities such as work-space logistics, material deliveries, and resource allocation. However, construction equipment management suffers from some deficiencies in different levels of site planning and their movements. Studies indicate that construction site planning highly depends on individuals' experiences, imaginations, and intuitions. Even though 4D visualization enhanced communications and mutual understanding of various stakeholders, there more pioneer computer-based technologies are employed to facilitate the visualization process. Immersive virtual technologies, such as VR technology, enable multiple design solutions to be evaluated digitally, which leads to is no effective tool to analyze, evaluate, and anticipate ongoing challenges and future difficulties [7]. Furthermore, in modular industrial projects, heavy mobile cranes are employed for on-site module installation at their set locations. Efficient planning is essential to decrease errors, accidents, additional costs, and schedule delays [8]. An effective crane planning in these kinds of projects includes many tasks (i.e. crane type selection, crane location optimization, lift sequencing, lift path planning). Heavy lift path planning is a pivotal subtask in heavy lift planning. Many scholars proposed computer-based solutions to optimize or enhance the process by automation [8]–[10].

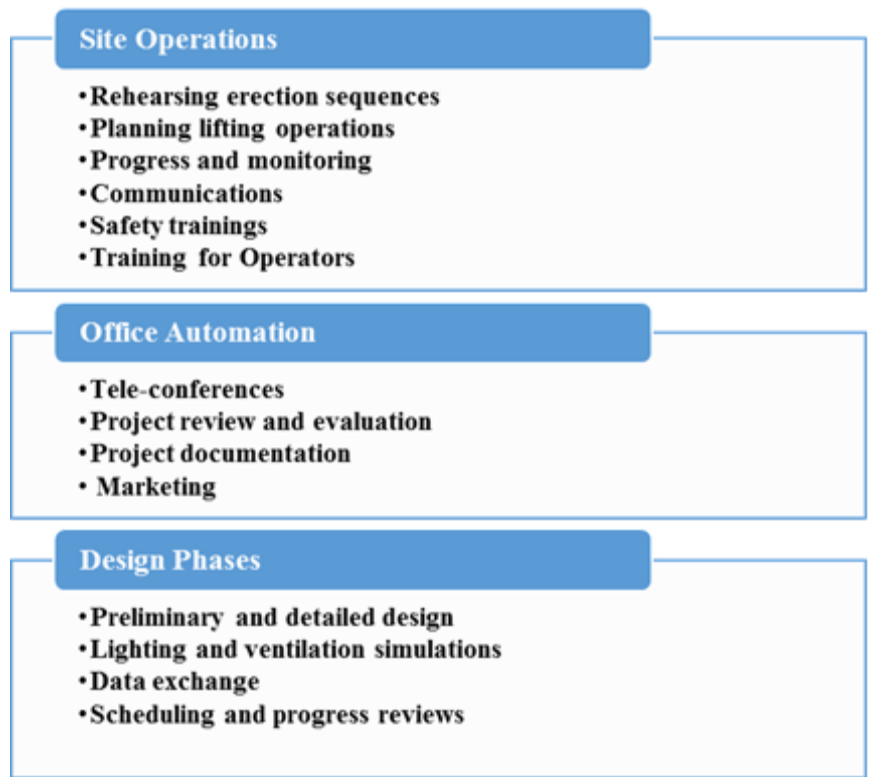

Figure 1- Potential Applications of VR in Construction [5]

In order to increase the construction productivity, investigations are conducted to link these novel visualization technologies to crane planning and management. Some scholars employed the VR technology to provide a virtual environment to train cranes operators with neither dangers or expenses. For instance, overhead crane [11] and bridge crane [12] operator training systems based on virtual reality are discussed. Some other efforts are made to evaluate and enhance safety conditions in this regard [13]–[15]. However, authors have not found any specific investigation in the application of VR technology in heavy lift planning process of mobile cranes.

## 3 Development of VR-Based Heavy Lift Planning Simulator

The objective of our approach is to unify a virtual environment for different use cases in lift planning and execution of heavy lifts of prefabricated modules in industrial projects. Using an interactable virtual environment (VE) would enable users to have more profound experiences. Virtual reality (VR) is a mixture of digital processing, computer graphics, multimedia technology, sensor technology, and other computer-based information technologies. Thanks to VR technology, users are able to freely navigate any simulated 3D models in a VE.

Moreover, they would be capable to actively interact with the VE. Navigation is one of these interactions. Navigation includes the ability to move around and explore the simulated objects in a virtual environment. Similarly, interaction is the ability to control and affect the VE and also be affected by it. For example, the ability to move objects in a VE is an interaction between the user and the VE. The most significant advantage of VR models to other forms of VE such as personal computers (PC) displays is providing a lifelike three-dimensional and 360° experience via a wearable display-headset with a higher degree of virtual immersion.

In this research, a prominent VR headset, Oculus Rift is employed which consists of several parts. A head-mounted display (HMD), an Xbox Controller, and Oculus Rift Constellation sensor, an optical-based tracking system developed by Oculus VR, are the main parts of the VR device. The VR device needs to be supported by a high-performance graphical process unit (GPU). Thus, the user can experience a more realistic environment than simple 3D-glasses.

This paper presents the overall concepts of VrCrane, a VR simulator for heavy mobile cranes (see Fig 2). Authors proceeded five main steps in order to achieve a VR tool which facilitates the lift planning for heavy mobile cranes. In this regard, a 3D model of the selected crane and the construction site is required. It is possible to use BIM models or a raw 3D model and a manual database. Hence, a pilot BIM model of an industrial plant is employed which has both information and 3D model. Moreover, it is required to make users able to control mobile cranes' every possible degree of freedom.

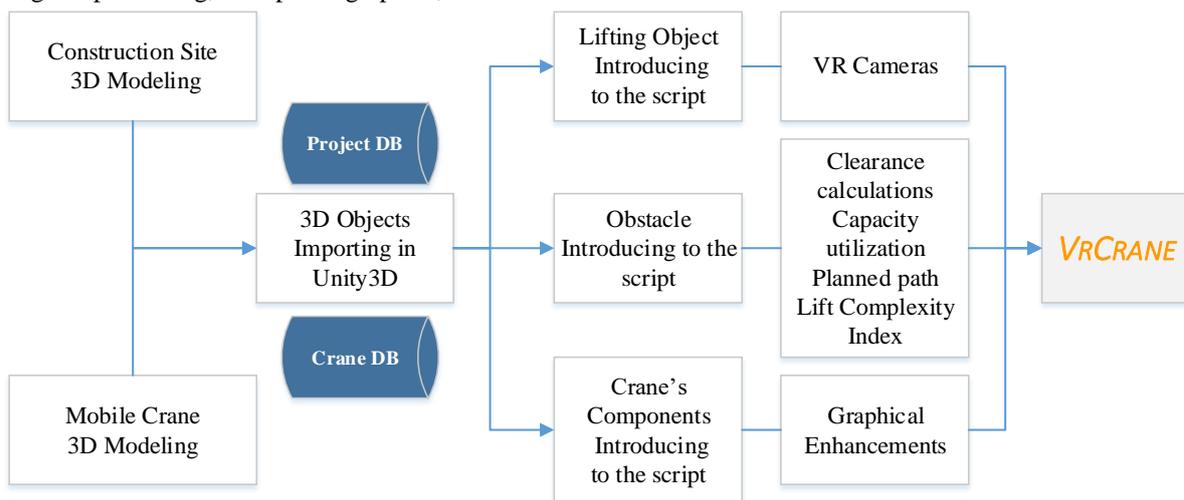

Figure 2 - VR-Crane Development System

Therefore, Uinty3D as a handy and free game engine is used to simulate its every six degrees of freedom. These degrees of freedom are controlled by the Xbox controller (see Fig 3). Various types of virtual cameras equip the simulated model to make users experience the lifting operation as an operator, a signalman, or through other views which is easily possible in a virtual environment (i.e. bird view, plan view, dynamic view, etc.). Extra real-time calculations and information presentations (i.e. capacity usage percentage, clearance color-coding, risk evaluations, etc.) for virtual operators and others would increase the common understanding and facilitate the communication process among all decision makers and other key stakeholders (Fig 4). In addition, a life-like experience of a heavy lift would reveal the underlying challenges which were previously concealed during typical 2D lift planning.

## 4 Conclusions

By the development of virtual environments such as virtual reality, augmented reality and mixed reality in architecture engineering and construction (AEC) industry, it is an indisputable need for stakeholders and planners to use a better approach for their labor-intensive and error-prone activities particularly in heavy lift planning. Provided a virtual reality environment in which the user can experience lifting process in an immerse virtual environment not only facilitates planning for critical lifts (e.g. modules, heavy vessels) but also provides a training environment to enhance safe climate prior to the actual lift in a safe, fast, and low-priced manner.

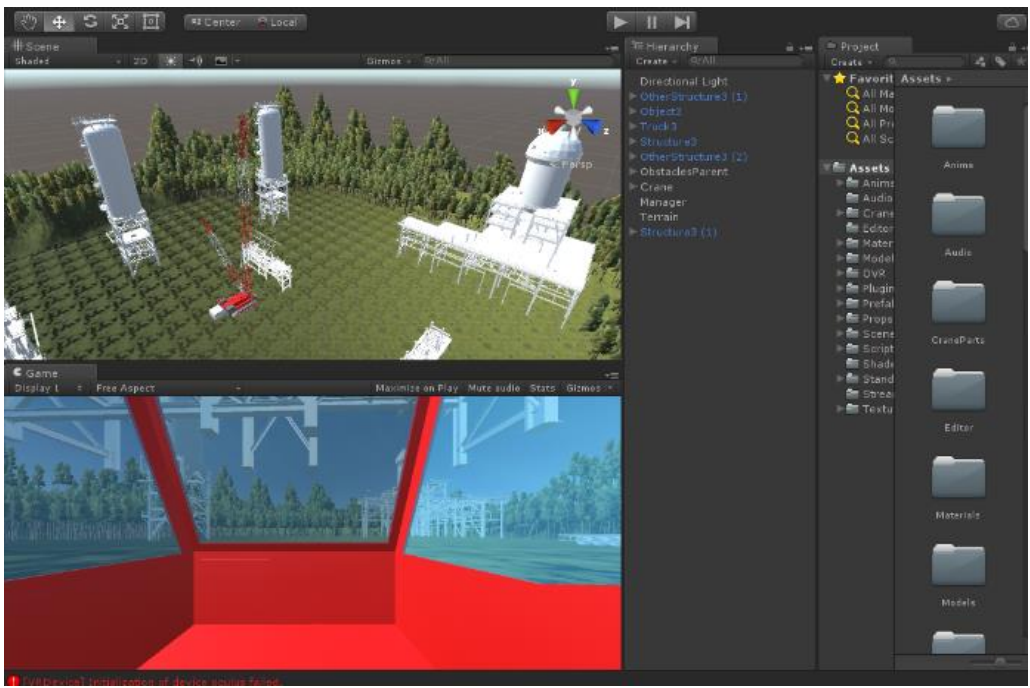

Figure 3- Simulation Environment in Unity3D

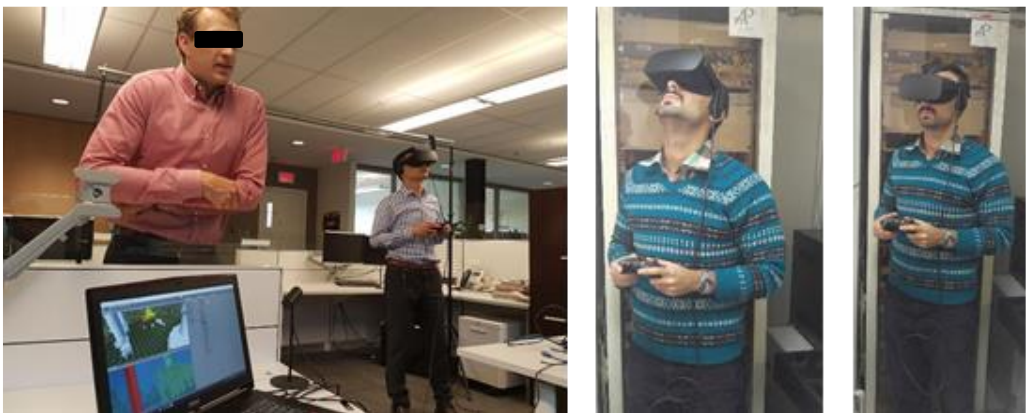

Figure 4- Users are experiencing the virtual lift operation

## 5  Acknowledgment

Sincere supports from PCL Industrial Management Inc. and TECNOSA R&D Center are greatly appreciated. The writers are also thankful to those who participated in this study and supported the research project.